\documentstyle[12pt]{article}

\input epsf

\relax

\begin{document}
\pagestyle{empty}
\begin{flushright}
{CERN-TH.6861/93}
\end{flushright}

\vspace*{5mm}

\begin{center}

{\large \bf COLLIDER JETS IN PERTURBATION THEORY$^{\dag}$}

\vspace*{1cm}

{\bf Stephen D.\ Ellis$^{\ddag}$} \\
\vspace{0.3cm}
Theoretical Physics Division, CERN \\
CH - 1211 Geneva 23 \\
Switzerland \\
\vspace*{2cm}
{\bf Abstract} \\ \end{center}
\vspace*{5mm}

\noindent
Recent progress in the {\em perturbative}
analysis of hadronic jets, especially in the context of $pp$ and $p
\overline{p}$ colliders, is discussed.
The characteristic feature of this work is the emergence
of a level of precision in the study of the strong interactions far beyond
that previously possible.
Inclusive cross sections for high energy jets at the
Tevatron are now perturbatively calculable with a reliability on the order of
10\%. At present this theoretical precision is comparable to
(if not somewhat better than) the quoted
experimental errors.  Progress has also been made towards understanding
both the internal structure of jets and the influence of the details
of the jet-defining algorithm.

\vspace*{5cm}
\begin{flushleft} CERN-TH.6861/93 \\
June 1993
\end{flushleft}

\vspace*{0.5cm} \noindent
\rule[.1in]{16.5cm}{.002in}

\noindent
$^{\dag)}$ Talk presented at the XXVIIIth Rencontres de Moriond,
{\it QCD and High Energy Hadronic Interactions}, March 1993.
\medskip

\noindent
$^{\ddag)}$ Permanent address: Department of Physics FM-15, University of
Washington, Seattle, WA, USA. \vspace*{0.5cm}

\vfill\eject

\setcounter{page}{1} \pagestyle{plain}

\section{Goals}

Why study jets? The basic goal is to be able to perform ``precision''
studies of the strong interactions, a prospect unthinkable just a few
years ago. In this context the term
``precision'' implies a theoretical uncertainty
no larger than 10\%.
Intuitively the study of jets is a natural arena for precision since the
jets can be thought of as the ``footprints'' of the underlying partons. It is
at the parton level that we have the best control of the strong interactions
through perturbative QCD.\footnote{
\setlength{\baselineskip}{ 10truept plus 0.2truept minus 0.2truept}
This can be (half-seriously) contrasted with studies performed
\baselineskip 10truept plus 0.2truept minus 0.2truept
with Monte Carlo simulations, where we note that one of the
\baselineskip 10truept plus 0.2truept minus 0.2truept
definitions \cite{dictionary} of the noun ``simulation'' is ``the assumption
\baselineskip 10truept plus 0.2truept minus 0.2truept
of a false appearance''.
\baselineskip 10truept plus 0.2truept minus 0.2truept}
This perturbative approach applies not only to
hadron-hadron collisions\cite{EKS,EKS2,greco,kosower}, as discussed here,
but also to $ep$ collisions\cite{ep} and $e^+ e^-$ collisions\cite{ee}.

With such a precise tool one can attack issues such as looking for
deviations from the Standard Model at very short distances due, for
example, to
compositeness. One can study jet production to learn more
about the structure of the hadrons, especially the gluon distribution
function. Finally one can study the jets in detail in order to
possibly differentiate
quark-initiated and gluon-initiated jets\cite{pump}
so as to control triggers and
backgrounds in the search for $p \overline{p} \to H + X \to
W^+ W^- (ZZ) +
X \to 4 $ jets $ + X^\prime$ at the SSC and LHC.

\section{What is a Jet?}

Clearly the first question that we must answer is how to define (and find)
jets. The qualitative goal is clear. The relatively isolated sprays of
energetic hadrons observed in the final states of high energy collisions are
naively (and correctly) associated with the production of isolated,
energetic partons via the
scattering of small numbers of partons. These large
angle scatterings involve only short distance interactions where the strong
interactions are relatively weak and perturbative techniques are
appropriate. Further, since we will deal with {\em inclusive} definitions of
features of the final state, one anticipates that nonperturbative corrections
to the perturbative results are small. However, although we would like to
associate a {\em unique}
subset of the final hadrons with {\em the }jet from a
single scattered parton, we know that such a mapping cannot, in principle,
be precise. The partons (quarks and gluons) carry color charge and are
(treated as) essentially massless in the theoretical calculation. On the
other hand a jet of hadrons has no color charge and often large invariant
mass. Jets must arise from the coherent, collaborative activities of
several ($\ge 2$) partons. Thus jets are necessarily somewhat ambiguous
objects and we wish to treat them
in such a way that these unavoidable ambiguities
do not play an important role.

We need a jet definition or algorithm that, while it is {\it a priori}
arbitrary, must still satisfy certain well defined constraints. It must be
reliable and easy to use for both theorists and experimentalists. For the
former this means ``infrared finite'' order by order in perturbation theory,
while the latter demand an algorithm that is straightforward,
efficient and well suited to the experimental situation. In the
$p\overline{p}$ case this means jets defined in terms of
the natural variables of
longitudinal phase space, the pseudo-rapidity $\eta =\ln (\cot [\theta /2])$,
the azimuthal angle $\phi $, and the ``transverse component'' of the
energy $E_T=E\sin \theta $. We choose to use the {\em cone} algorithm
outlined in the so-called ``Snowmass Accord''\cite{snowmass}. To a good
approximation this algorithm is employed by the CDF Collaboration\cite
{CDFlong}, although we will see below that there are still some issues to be
resolved~in practice.

The jet from the cone algorithm is typically defined in terms
of the particles $n$ whose momenta $\overrightarrow{p_n}$ lie within a cone
centered on the jet axis ($\eta _J,\phi _J$) in pseudo-rapidity $\eta $ and
azimuthal angle $\phi $,
$\sqrt{(\eta _n-\eta _J)^2+(\phi _n-\phi _J)^2}<R\,$.
The jet angles ($\eta _J,\phi _J$) are the averages of the particles'
angles,
\begin{equation}
\label{snow}
   \eta_J   =  \sum_{n\in {\rm cone}} p_{T,n} \eta_n/E_{T,J} ,
\quad
   \phi_J   =  \sum_{n\in {\rm cone}} p_{T,n} \phi_n/E_{T,J} \,
\quad ; \quad
E_{T,J}=\sum_{n\in {\rm cone}}p_{T,n}\,.
\end{equation}
This process of finding the center and then recalculating the cone
is iterated until the cone center matches the jet center ($\eta _J,\phi _J$).

This definition implies that, for a given hadronic final state, one
identifies all jets that satisfy the algorithm (typically with jet $E_{T,J}$
above some lower threshold $E_{T,{\rm min}}$). This process can and does
lead to situations where individual hadrons are members of more than one
jet, i.e. the cones are found to overlap. Hence the cone algorithm must be
augmented to handle this situation and we will return to this point below.
It will not have a large numerical
impact on the jet cross section but will effect
the observed internal structure of jets.

\section{Uncertainties}

Recall that the general theoretical structure of the perturbative jet cross
section calculation for $A+B\to$ jet $+X$ has the schematic form
\begin{eqnarray}
\label{form}
{\frac{{d\sigma }}{{dE_Td\eta }}}
(\eta ;E_T,s;\Lambda _{QCD};\mu;R) &\sim&
\int \int dx_adx_b\Pi _idk_i
G_{a/A}(x_a;\Lambda _{QCD};\mu) G_{b/B}(x_b;\Lambda _{QCD};\mu)
\nonumber \\
&\times&
{\frac{{d{\hat \sigma }}}{{dk_i}}}(a+b\to i;x_a,x_b,s,k_i;\Lambda _{QCD};\mu)
\nonumber \\
&\times& S_{jet}(k_i,E_T,\eta ;R) \  .
\end{eqnarray}
The various components include the parton distribution functions,
$G_{i/I}(x_i;\Lambda _{QCD};\mu )$,
the parton-parton scattering cross section,
$d{\hat \sigma}
/dk_i$, and the jet algorithm or ``projection'' function $S_{jet}$ that
identifies the jet in the final state partons. Note the explicit dependence
on the {\it a priori} arbitrary theoretical parameter $\mu $, the
factorization/renormalization scale, which is an unavoidable feature of
finite order perturbation theory. The inclusive jet cross
section evaluated\cite{EKS2} at order $\alpha _s^3$ is compared
with CDF data\cite{CDFjet} in Fig.~1a. Note the outstanding
agreement between theory and data over nearly ten orders of magnitude. These
results are for $\sqrt{s}=1800$ GeV and are averaged over the rapidity range
$0.1\leq |\eta |\leq 0.7$ with a cone of size $R=0.7$. The theoretical
parameter $\mu $ is set to $E_T/2$ as discussed below and the HMRS(B) parton
distribution functions\cite{HMRS} are employed. This comparison is pursued
in more detail in Fig.~1b, where the difference between the data and the
theory, scaled by the theory, is exhibited (the reference
theoretical result is the dotted
line). Unlike Fig.~1a where the full experimental
systematic errors are indicated, in Fig.~1b the error
bars include only the $E_T$ dependent systematic uncertainties while the $E_T$
independent uncertainty is indicated by the dashed lines ($\sim \pm 20 $\%).
The
curves correspond to $\mu = E_T/4$ (solid)
and $\mu = E_T$ (dashed). (The dot-dashed curve will be
discussed below.)  It is, in fact, difficult to distinguish the various
theoretical curves in the figure
and this feature is precisely the point.

 From the consideration of Fig.~1b
we learn that the absolute agreement between data and theory, at
least for $E_T >$ 50 GeV, satisfies our 10\% goal (actually better
agreement than required by the stated
overall normalization uncertainty in the data
of order 20\%). We also see that the
theoretical result is reassuringly independent of the arbitrary scale $\mu $,
varying by only 10\% in the ``physically relevant''
range $E_T>\mu >E_T/4$. Note that both of these limit values
yield cross sections {\em below}
that for $E_T/2$, which is a local extremum.  If
we accept this stability in $\mu $ as a reliable measure of the
desired 10\% theoretical
uncertainty, we learn the important fact that the reliability of the order $%
\alpha _s^3$ degrades dramatically for $E_T$ below 50 GeV as indicated by
the divergence of the solid and dashed curves in this regime.
It is troubling to note
that this behavior essentially scales in $x_T=2E_T/\sqrt{s}$. Thus this
order in perturbation theory is not to be considered reliable (at the 10\%
level) to describe jet physics at the SSC or LHC at $E_T$ values much below
1000 GeV!

The parton distribution functions,
$G_{i/I}(x_i;\Lambda _{QCD};\mu )$, especially the gluon
component, are not precisely known and until recently were a ``major'' source
of uncertainty at the level of 20\%. However, the more recent
fits\cite{mrs,cteq}
exhibit jet cross sections with differences at the 10\% level
(and good agreement with the HMRS(B) distributions used here).

Since Eq.~(\ref{form}) represents a purely perturbative result, it contains
no explicit nonperturbative effects, either from fragmentation smearing or
from the underlying event. The former effect, which is intended to
characterize how the partons interact coherently to form the final
color singlet hadrons, is thought to involve some amount of
momentum transfer transverse to the original parton direction. Thus, while
the process must conserve overall $E$ and $\overrightarrow{p}$, the $E_T$ of
the final hadrons can be
somewhat redirected from that of the partons. If the
characteristic momentum transfer is of order 0.5 GeV, this effect should
result in a smearing of angles of the magnitude
$\Delta \Theta \leq 0.5$ GeV$/\left\langle E_{\rm Hadron}\right\rangle$.
This will be unimportant for energetic hadrons ($E_{\rm {Hadron}}> 10$
GeV) and reasonably sized jets ($0.4 < R < 1.0$). The second and, perhaps, more
important effect not included in the calculations is
the underlying event. This term is intended to describe the soft interactions
of the remaining partons in the initial hadrons that, while not
participating in the short-distance, large $p_T$ interaction described by
Eq.~(\ref{form}), still can contribute final hadrons and thus $E_T$ to the
jet of interest. To the extent that such soft interactions generate a
a fairly uniform distribution of particles in the variables $\eta $ and $%
\phi $, as in usual minimum bias events, the contribution to the jet is
essentially a geometric effect. This is one of the advantages of
such a cone definition for the jet. With an observed $E_T$ density,
in $\eta,\phi $ units, of order 1~GeV (see below) in minimum bias events,
jet sizes
characterized by $\pi R^2 \sim \pi \left( 0.7\right) ^2 \sim
1.5$ and a logarithmic derivative for the differential cross section with
respect to $E_T$ of order 6 (i.e. $d\sigma /dE_T\propto E_T^{-6}$ in the
range of interest) we have
${\Delta \sigma }/\sigma \leq 9$ GeV$/{E_T}\leq 10\% $
for $E_T>100$ GeV.  This is
essentially the same range of reliability as defined by the
perturbative effects
discussed earlier. We will return to the issue of the underlying
event later.

Finally there is expected to be some uncertainty involving the specific jet
algorithm, $S_{jet}$, itself.  While the ``Snowmass Accord'' was intended to
constitute an agreement by both theorists and experimentalists to use the
identical jet algorithm, in the event, ``real-life physics'' is somewhat
more complicated. In particular the ``Snowmass Accord'' does not treat the
issue of how overlapping jets identified by the algorithm are ``merged'' and
we will discuss this point below. However, it is important to recognize that
the effect of this issue on the inclusive jet cross section is numerically
small, within our working limit of 10\% (see the dot-dashed curve
in Fig.~1b, which is explained below).

\section{Jet Shape Dependence and the Jet $E_T$ Profile}

Coupled to the increased numerical reliability of the order $\alpha _s^3$
cross section, is the attractive feature that, like the data, the
theoretical result
depends on the specific details of the jet algorithm. In the present context
this implies a dependence on the jet ``size'' parameter $R$. While such
dependence appears only at ``lowest order'', in some sense,
in the $\alpha _s^3$
calculation, it is of interest to compare the dependence with that observed
in the data.\cite{CDFjet}  In a careful study of this issue in
Ref.~\cite{EKS} two features stand out.
First the region of stability in $\mu $, as defined above, is also
correlated with the $R$ dependence. At both large $R$ ($R>1$) and small $R$
($R<0.4$), the theoretical cross section exhibits a monotonic dependence
on $\mu $, reminiscent of the Born cross section. This feature is indicated by
the solid and dashed curves in Fig.~2a showing the inclusive jet cross
section versus $R$ with various $\mu$ values (solid = $E_T/2$, long dashed =
$E_T/4$, short dashed = $E_T$,
the dot-dashed curve
will be explained below).
Note that the three curves intersect in the region of $R \sim 0.7$ but that the
cross section increases monotonically with $\mu$ at small $R$ and decreases
at large $R$.
Thus, at least at this order
of perturbation theory, the theory itself gives a hint as to the optimum
value of $R$ at which to compare theory and data, $R\sim 0.7$. Luckily this
is just where CDF has been working!

The second point contained in Ref.~\cite{EKS} involves the actual dependence
on $R$ of the cross section.  As indicated in Fig.~2a the reference value
$\mu = E_T/2$ yields a cross section that
does not vary as rapidly with $R$ as the data of CDF.  In some crude sense
the data are suggesting the need for ``fatter'' jets.  Changing to a smaller
$\mu$ value (here $\mu = E_T/4$) leads to a larger $\alpha_s$ value,
more radiated gluons, ``fatter''
jets and more rapid $R$ variation.

The $E_T$ distribution within
the cone of the jet can be analysed more directly by studying the fractional
$E_T$ profile, $F(r,R,E_T)$.
Given a sample of jets of transverse energy $E_T$ defined with a cone radius
$R$, $F(r,R,E_T)$ is the average fraction of the jets' transverse energy
that lies inside an inner cone of radius $r<R$ (concentric with the
jet-defining cone). Thus the quantity $1-F(r,R,E_T)$ describes the
fraction of $E_T$ that lies in the annulus between $r$ and $R$.
It is this quantity that is most easily calculated in perturbation theory
as it
avoids the collinear singularities at $r=0$.
The results for $F$ with the three $\mu$ values are plotted in Fig.~2b
versus the inner radius $r$ with $R=1.0$
for $E_T=100$ GeV and compared to CDF data.\cite{CDFcone} (The dot-dashed
curve will be explained below.) As with the $R$
dependence discussed above, $F$ is being calculated to
lowest nontrivial order and thus exhibits monotonic $\mu $ dependence. While
there is crude agreement between theory and experiment, the theory curves
are systematically below the data. This
situation suggests that the theoretical jets have too large a fraction of
their $E_T$ near the edge of the jet ($r\simeq R$).

We have seen that the
$R$ dependence of the cross section suggests
the importance of higher-order contributions
to increase the level of associated
radiation, at least near the center of the cone. At the same time our
considerations of $F$ suggest that the
data favor a reduction of the $E_T$ fraction near the edge of the cone.
Although these conclusions seem initially to be
contradictory, the likely consistent
explanation is based on a
detailed but important physical point concerning
how the jets are defined. The issue, as mentioned earlier, is that of {\em
merging}, i.e.
how close in angle should two partons be in order to be associated
as a single jet. In a real experiment such a situation is presumably
realized as two sprays of hadrons, each with a finite angular size
that arises from
both fragmentation effects and real experimental angular resolution effects.
If the angular separation is large enough, there is a valley in the $E_T$
distribution between the two sprays and experimental jet-finding algorithms
will tend to recognize this situation as two distinct jets. Recall that we
expect for jets of $E_T>100$ GeV that the angular extent of fragmentation
effects will be small compared to the defined jet cone sizes. However, the
theoretical jet algorithm defined in strict adherence to the
``Snowmass Accord'' will merge two partons into a single
jet whenever it is mathematically possible. Thus the limiting
configuration with two equal transverse energy partons $2R$ apart will
be counted as a single jet with its cone centered in the ``valley''
between the two partons.
A ``real experimental'' treatment of this configuration
is unlikely to identify it
as a single jet.
To simulate the experimental
algorithm in a simple way we add an extra constraint
in our theoretical jet algorithm. When two partons, $a$ and $b,$ are separated
by more than $R_{sep}$, $R_{ab}=\left[ (\eta _a-\eta
_b^{})^2+(\phi _a-\phi _b)^2\right] ^{1/2}\geq R_{sep} (\leq 2R)$,
we no longer merge
them into a single jet.  As an example, the results of
calculating both the $R$ dependence and the $E_T$ fraction $F$ with
$R_{sep}=1.3R$ and $\mu =E_T/4$ are illustrated by the dot-dashed curves in
Figs.~2a and b. Clearly the extra constraint of $R_{sep}$ has ensured that
there is approximately the observed fraction of $E_T$ near the edge of the
cone while the reduced $\mu $ value has increased the amount of associated
radiation near the center of the cone and produced a larger variation with
$R$. The specific choice $R_{sep}=1.3R$ is also in good agreement with the
detailed CDF study\cite{CDFlong} of this issue.
It is important to note that, while these limiting configurations of
the partons make important contributions to $F$ for $r \sim R$, they
constitute only a small contribution to the jet cross section itself.
Hence the cross section is
relatively insensitive to the
parameter $R_{sep}$, decreasing by $\leq $ 10\% as $R_{sep}$ is
reduced from $2R$ to $1.3R$ with fixed $\mu $ for $E_T=100$ GeV.
This point is illustrated in Fig.~1b by the dot-dashed curve that
indicates the small change in the cross section from the
reference result.

\section{Scaling in Jet Cross Sections}

One of the most interesting new measurements from the CDF
Collaboration\cite{CDFscale} involves the comparison of jet cross sections at
two different center-of-mass energies, $\sqrt{s} = 1800$ and 546 GeV.
By comparing the two cross sections, multiplied by $E_T^3$ to obtain
dimensionless quantities, at fixed $x_T$ values one has a check on
the scaling violation as predicted by QCD.  The resulting ratio is
illustrated in Fig.~3.  The data are clearly inconsistent
not only with pure scaling
(a ratio of unity) but also with the cross section calculations we have
discussed up to now.  It is important to note that the data and theory
at the two different
energies agree within the full systematic errors
(as we saw in detail above for the 1800 GeV case) but
in the ratio much of the
systematic uncertainty cancels and the deviation
between theory and data appears to be of order 2$\sigma$
at low $x_T$. Note that the region of comparison,
$ 0.1 < x_T < 0.3$, is where
the theory was argued above to be reliable to 10\%.  Also
note that, while the individual cross sections
are sensitive to the specific choice of parton distribution functions,
the ratio is remarkably {\em insensitive} to that
choice.  This point is illustrated by the dashed curve in Fig.~3
corresponding to the
quite different parton distributions constructed by Berger
and Meng\cite{BM}.
While there is evidently not a large deviation between data and theory,
I suspect that at least one
contributing issue has physics interest and I will discuss it briefly
here.

The important point is how the data are corrected for the contribution
from the underlying event.  As suggested earlier it is presumably a good
approximation to treat the underlying event as an essentially uniform
(in $\eta$
and $\phi$) distribution of $E_T$ with only minimal correlation with the
hard scattering process
(for earlier discussion of this issue, both experimental and theoretical,
see Refs.~\cite{UA1} and \cite{marweb}).  This ``splash-in'' effect
will contribute to the jet $E_T$
simply on the basis of geometry. The issue here is then to evaluate the
level of this background contribution.  In the correction of the CDF
data\cite{CDFscale}
this underlying level is determined at each $\sqrt{s}$ by measuring the
$E_T$ density in jet events but at $90^\circ$ in $\phi$ to the observed
jet direction, the ``pedestal'' of the jet.
This yields an $E_T$ density of about 1.6 GeV/$R^2$
(1.0 GeV/$R^2$) at $\sqrt{s}=1800$ GeV (546 GeV).  This value is to be
contrasted with the corresponding $E_T$ density observed in minimum
bias events and found to be approximately 0.7 GeV/$R^2$ (0.5 GeV/$R^2$).
Although there is some selection bias in the jet trigger process towards
underlying events with larger than average local $E_T$ densities, the
suggestion here is that what is seen at $90^\circ$ to the jet can be thought
of as a fairly standard minimum bias event plus the contribution
of bremsstrahlung explicitly associated with the hard scattering, i.e.
associated ``splash-out''.  This latter contribution
is meant to be accounted for in the theoretical calculation (the $E_T$ density
observed {\em outside} of the jet cone in the theoretical calculation
is about half of the pedestal height observed experimentally in jet
events).  Thus to compare theory and data in first approximation,
the experimental jet $E_T$
should be corrected for a contribution from an underlying event just equal
to a minimum bias event.  This conclusion implies that the current data
have been {\em over}-corrected.  A ``plausible'' scenario is that the jet
$E_T$ in the 1800 GeV data have been over-corrected by 1.25 GeV
($\sim$(1.6 GeV $-$ 0.7 GeV)$\times \pi R^2$) while
for those at 546 GeV the over-correction is 1.0 GeV
($\sim$(1.0 GeV $-$ 0.5 GeV)$\times \pi R^2$
plus 0.25 GeV to account for non-perturbative splash-out effects being more
important at the lower $E_T$ values).  Instead of removing this correction
from the data it is easier (for me) to add it to the theoretical calculation.
The result is the dot-dashed curve
in Fig.~3 where we see that the disagreement
is now at about the 1$\sigma$ level.  While this improvement is perhaps not
overwhelming, it is relevant and I believe that the physics issues involved
are now more correctly treated, i.e. the underlying event contribution
is in reality more like a minimum bias event than the pedestal observed in
jet events.  (Note that this difference between these two scenarios
for the corrections
is meant to be spanned by the CDF error bars\cite{CDFscale}
and this feature presumably explains
why switching the central value from one to the other
improves the agreement by approximately 1$\sigma$.)

\section{New Jet Algorithms}

With no extra space left (either in the talk or this contribution), I will
just note that there has been recent work\cite{mike,ES} on the question of
replacing
the cone algorithm with
$e^+ e^-$ successive combination style algorithms for the study of jets
in hadron collisions.  While there are positive indications of qualitative
improvement
(e.g. the merging issue {\it per se} is removed), it is not yet clear
whether there is a quantitative improvement.

\section{Summary}

Let us briefly summarize what (I hope) we have learned.

\begin{itemize}

\item The theoretical calculations of one (and two) jet cross sections
at order $\alpha_s^3$ in perturbation theory are
reliable at essentially the 10\% level and now allow very precise comparisons
with data.  These comparisons enhance our confidence in perturbative QCD in
the large $E_T$ regime.
Unfortunately the calculations suggest much larger higher order contributions
for $x_T \leq 0.05$ corresponding to $E_T$ as large as 1 TeV at the SSC
(0.4 TeV at the LHC)!

\item At this order in perturbation theory the results are most reliable
for cone sizes around $R\sim 0.7$.  For cones sizes much smaller or
much larger than 0.7 higher orders must play an important role.

\item The analyses of the $R$ dependence of the cross section
and the $E_T$ profile $F$ yield
an even more detailed understanding of the structure of jets
and suggest that the ``merging'' issue must be taken into account.
These studies may also
yield an avenue for attacking the problem of differentiating
quark-initiated jets from gluon-initiated jets.

\item Further work is urgently required on the issues of scaling violations
and the role of the underlying event; of order $\alpha_s^4$ contributions for
the study of $E_T \sim M_W$ scale jets at the SSC/LHC; of $e^+ e^-$ jet studies
with cone algorithms to check the role of the underlying event; of studies of
$e^+ e^-$ style successive combination jet algorithms at hadron
colliders.

\end{itemize}

\newpage
\noindent
{\bf Acknowledgements}

It is a pleasure to acknowledge the essential contributions of my two
collaborators, Z.~Kunszt and D.E.~Soper,
to much of the work discussed here.  A special thanks also
goes to the CDF QCD group, especially J.~Huth, N.~Wainer and S.~Behrends,
for repeatedly and patiently explaining their measurements to me.
This work was supported in part by the
U.S. Department of Energy under grant DE-FG06-91ER-40614
and by the CERN TH Division, who are warmly thanked for their hospitality.

\bigskip\

\noindent
{\bf \Large Figure captions}

\begin{itemize}

\item[Figure 1:]
a) Inclusive jet cross section versus $E_T$ at
$\protect\sqrt{s}=1800$
GeV with $R=0.7$ comparing data\protect\cite{CDFjet}
with the order $\alpha_s^3$ result for
HMRS(B) parton distributions with $\mu = E_T/2$;
b) Scaled difference from theory result in
(a) comparing data and theory with $\mu = E_T/2$ (dots),
$\mu=E_T/4$ (solid),
$\mu=E_T$ (dashes), and $\mu=E_T/4,
R_{sep} = 1.3R$ (dot-dashed, as explained in the text).

\medskip

\item[Figure 2:]
a) Inclusive jet cross section data\protect\cite{CDFjet}
versus the cone size $R$ at $E_T=100$ GeV and
$\protect\sqrt{s}=1800$ GeV compared with
the standard order $\alpha_s^3$ result for
HMRS(B) parton distributions with $\mu = E_T/2$ (solid),
$\mu = E_T/4$ (long
dashed), $\mu = E_T$ (short dashed) and also with $\mu=E_T/4$,
$R_{sep}=1.3R$ (dot-dashed, as explained in the text); b)~$E_T$ fraction
$F(r,R,E_T)$ versus the inner radius $r$ for $R=1.0$, $E_T=100$ GeV and
$\protect\sqrt{s}=1800$ GeV with the four curves defined as in (a).

\medskip

\item[Figure 3:]
Ratio of the scaled cross
sections at $\protect\sqrt{s}=$ 546 and 1800 GeV
versus $x_T$ showing the CDF data\protect\cite{CDFscale} (with
the overall systematic uncertainty suggested by the dotted box)
and the theoretical results
for $\mu=E_T/4$ and $R_{sep}=1.3R$ with HMRS(B) parton distributions
(solid curve), Berger-Meng(A) parton
distributions\protect\cite{BM} (dashed curve) and the
situation correcting for the underlying event contribution as discussed in
the text (dot-dashed curve).

\end{itemize}

\newpage

\begin{figure}[h]
\centerline
    { \epsfbox{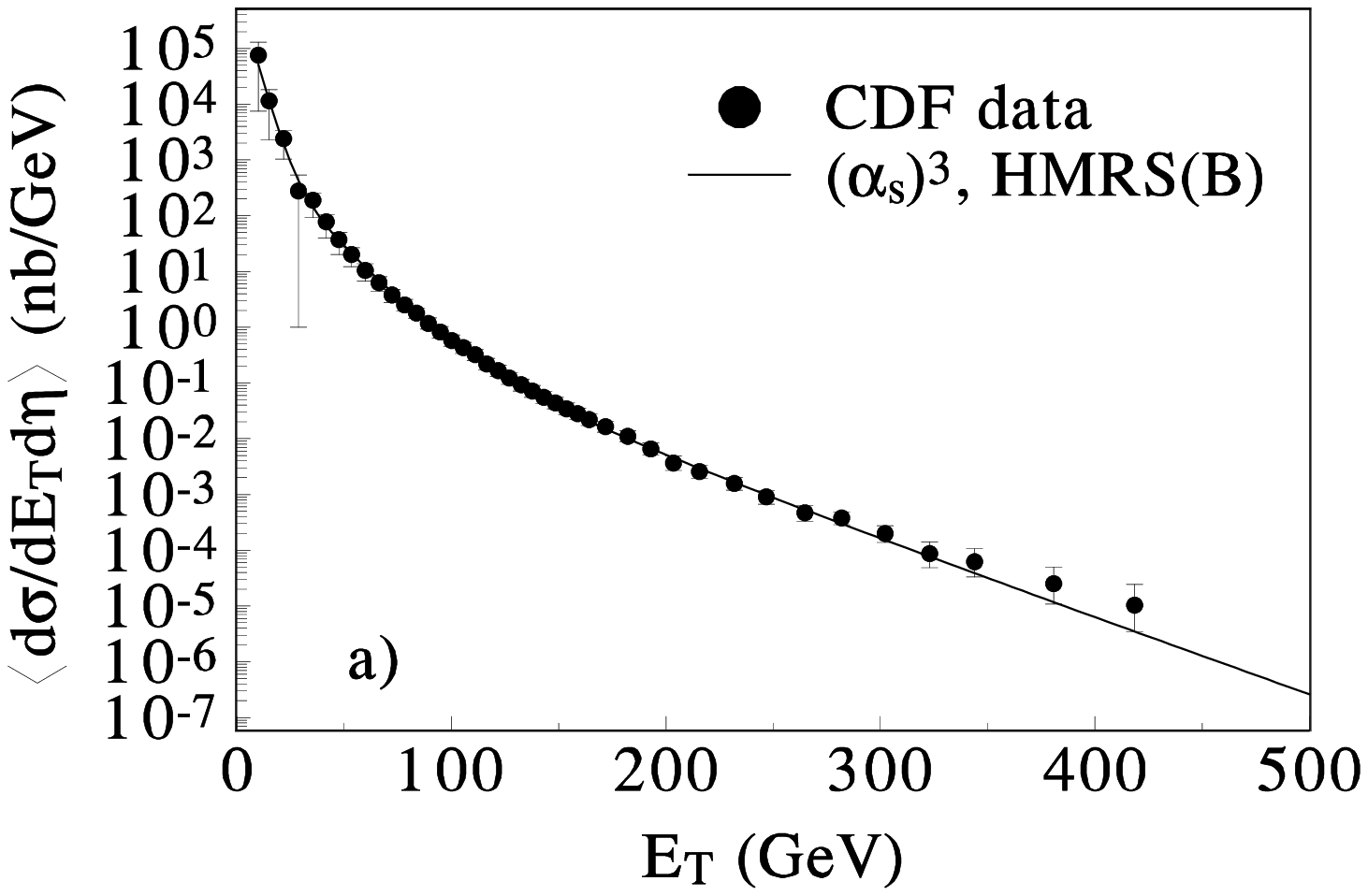} }

\vskip -0.9in

\centerline
    { \epsfbox{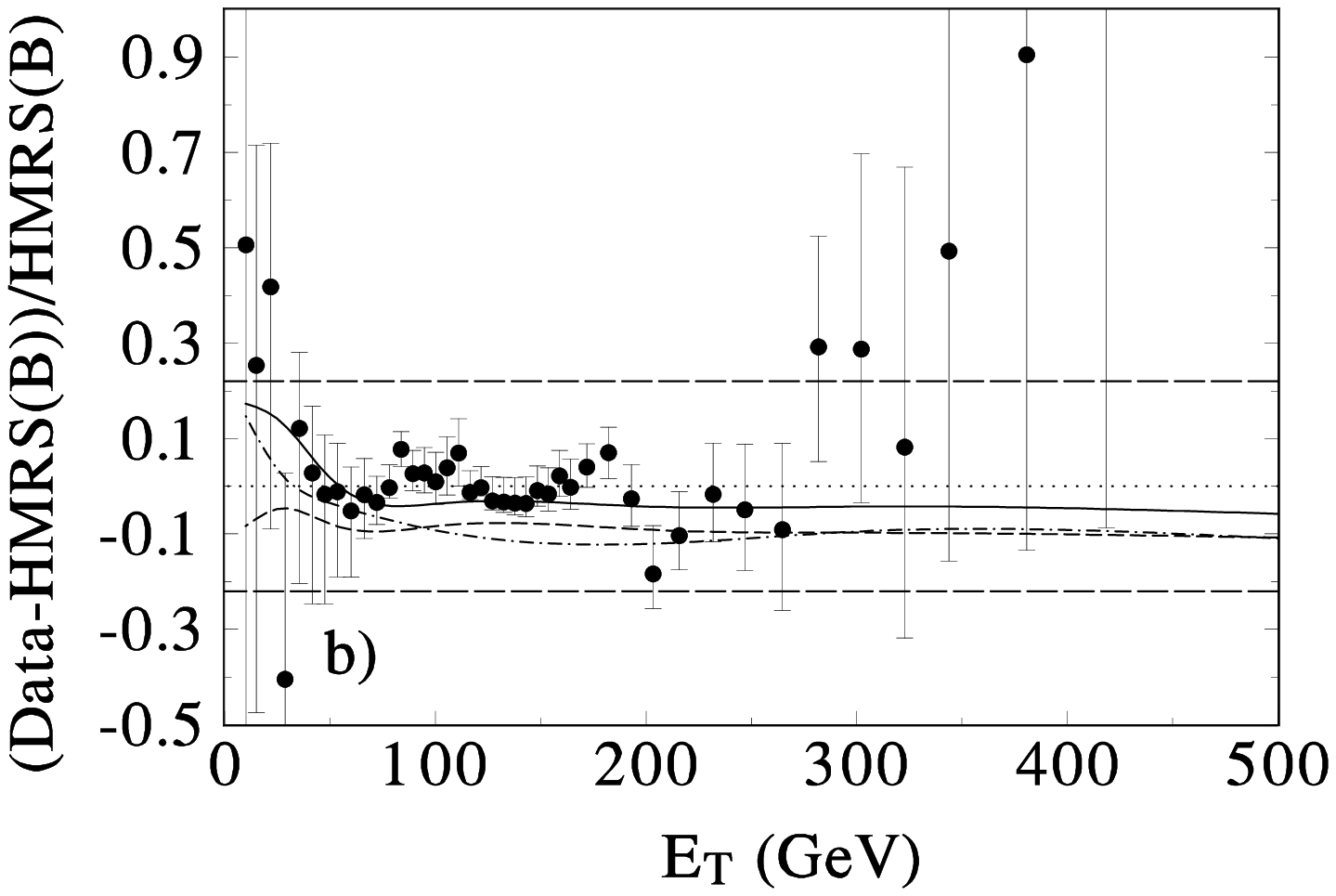} }

\centerline{Figure 1}

\end{figure}

\newpage

\begin{figure}[h]
\centerline
    { \epsfbox{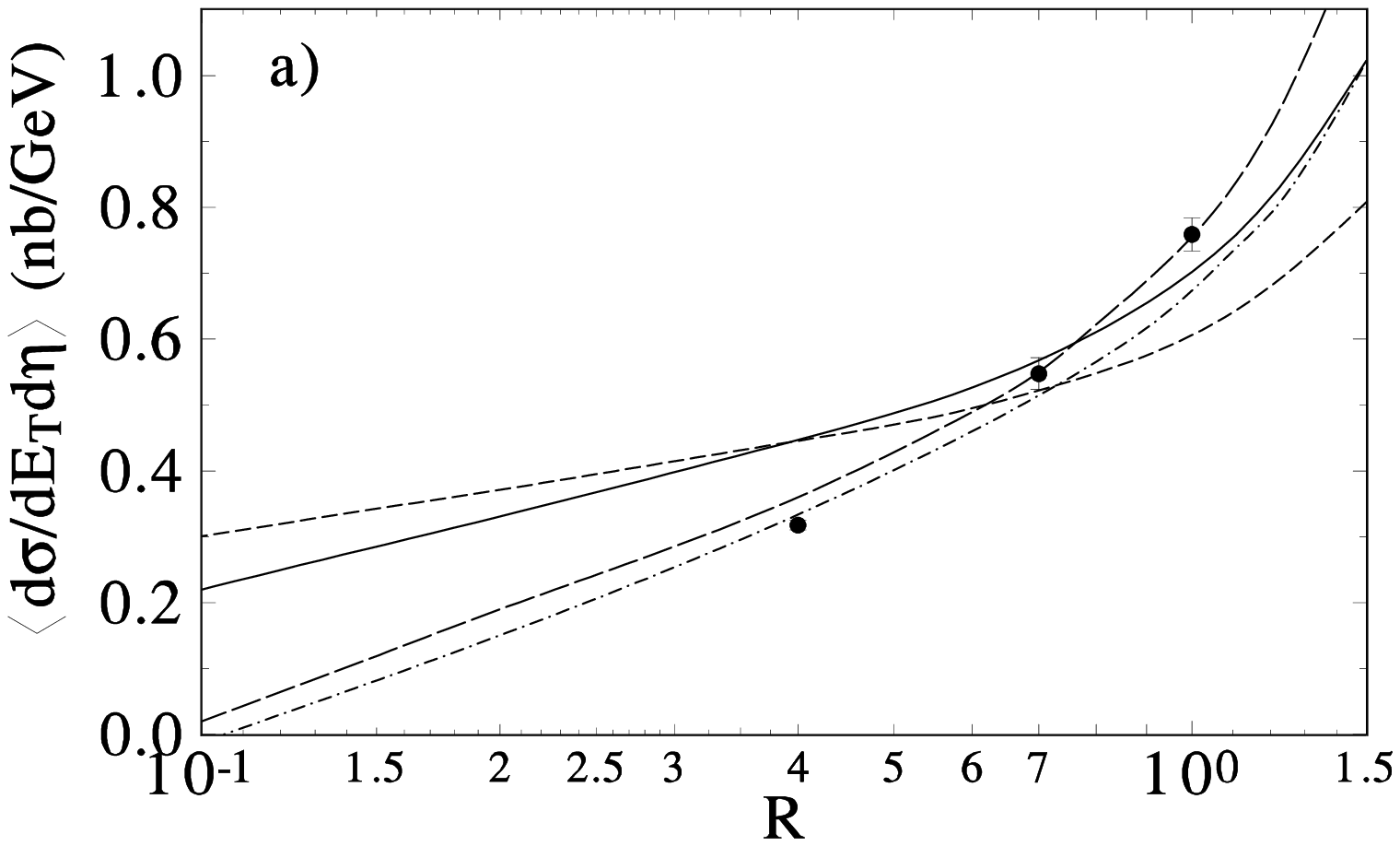} }

\vskip -0.9in

\centerline
    { \epsfbox{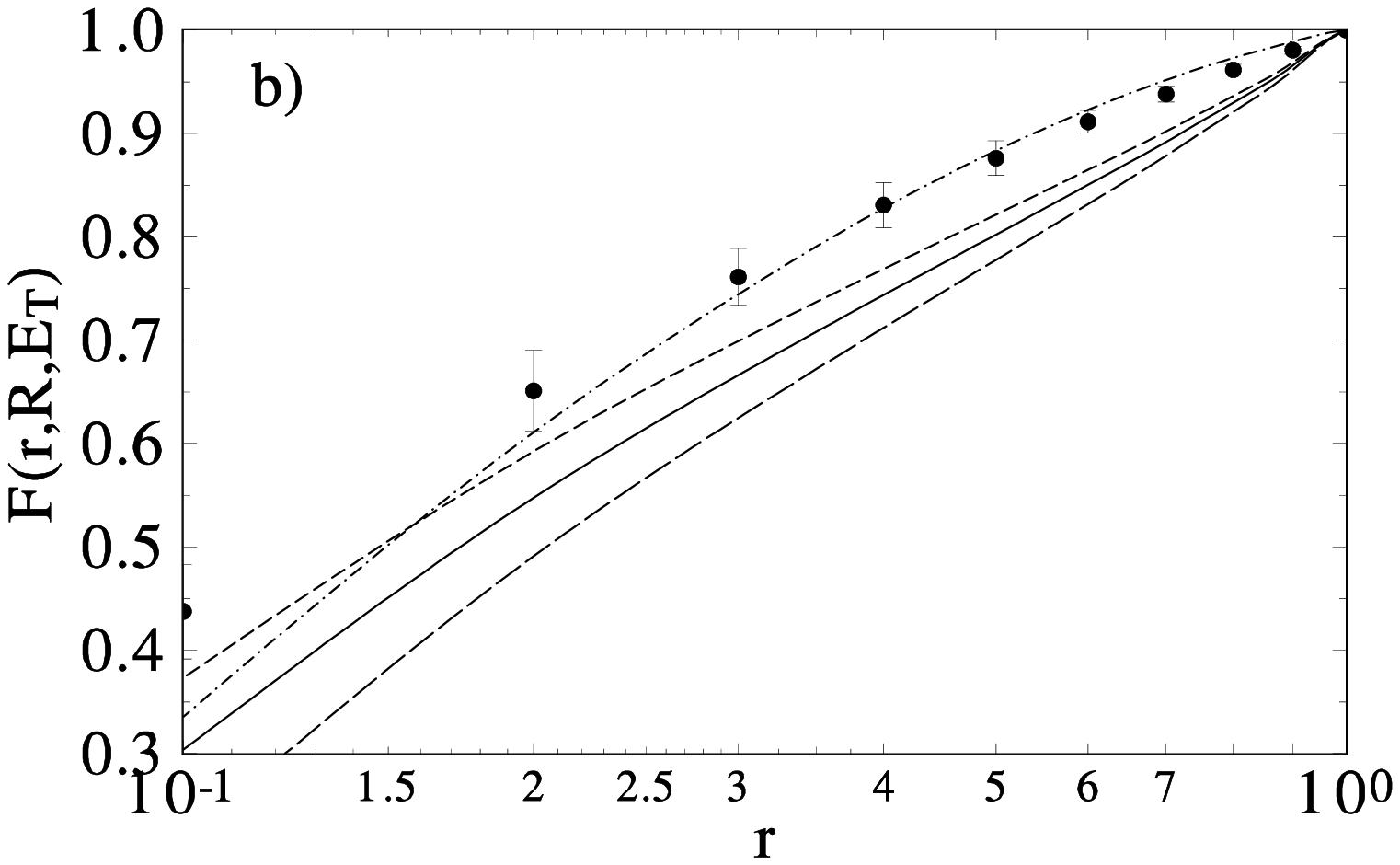} }

\centerline{Figure 2}

\end{figure}

\newpage

\begin{figure}[h]
\centerline
    { \epsfbox{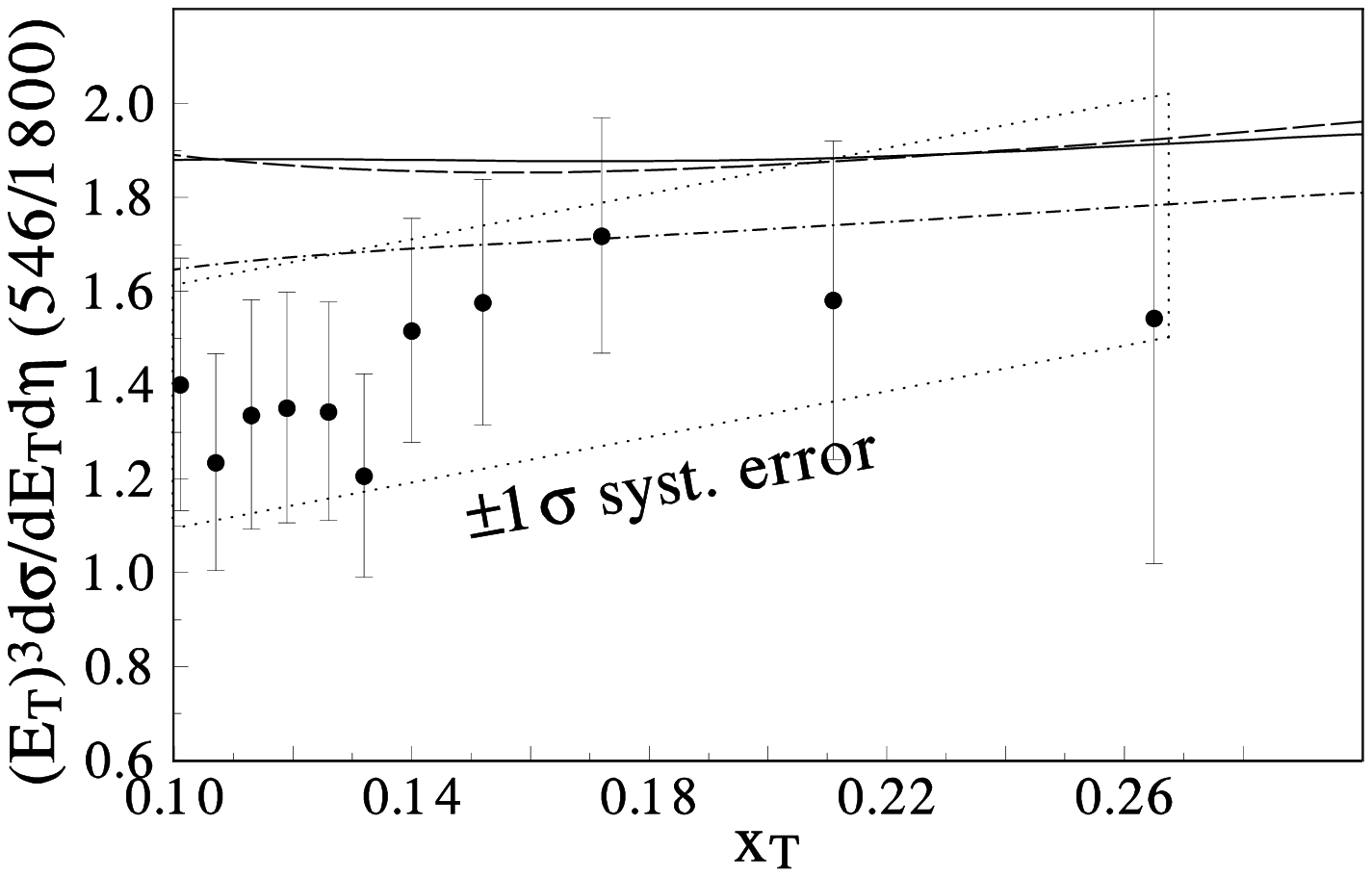}}

\centerline{Figure 3}

\end{figure}

\end{document}